\RequirePackage{fix-cm}
\let\accentvec\vec
\documentclass[twocolumn,final]{svjour3}

\let\vec\accentvec

\smartqed  % flush right qed marks, e.g. at end of proof
\usepackage{graphicx}
\usepackage{units}
\usepackage{amssymb,amsmath}
\usepackage{color}

\journalname{Appl. Phys. B}

\begin{document}

\title{A Nanofiber-Based Optical Conveyor Belt for Cold Atoms}
\titlerunning{A Nanofiber-Based Optical Conveyor Belt for Cold Atoms}

\author{Philipp Schneeweiss \and Samuel T. Dawkins \and Rudolf Mitsch \and Daniel Reitz \and Eugen Vetsch and Arno Rauschenbeutel*\thanks{* Arno.Rauschenbeutel@ati.ac.at}}
%\authorrunning{Short form of author list} % if too long for running head

\institute{P. Schneeweiss \and R. Mitsch \and D. Reitz \and A. Rauschenbeutel \at
              Vienna Center for Quantum Science and Technology,\\ TU Wien - Atominstitut, Stadionallee 2, 1020 Wien, Austria
%             \emph{Present address:} of F. Author  %  if needed
           \and
           S. T. Dawkins \and R. Mitsch \and D. Reitz \and E. Vetsch \ and A. Rauschenbeutel \at
              Institut f\"ur Physik, Johannes Gutenberg-Universit\"at Mainz, 55099 Mainz, Germany
}

\date{Received: date / Revised version: date}

%----------------------------------------------------------------------------------------------------------
\maketitle
\begin{abstract}
We demonstrate optical transport of cold cesium atoms over millimeter-scale distances along an optical nanofiber. The atoms are trapped in a one-dimensional optical lattice formed by a two-color evanescent field surrounding the nanofiber, far red- and blue-detuned with respect to the atomic transition. The blue-detuned field is a propagating nanofiber-guided mode while the red-detuned field is a standing-wave mode which leads to the periodic axial confinement of the atoms. Here, this standing wave is used for transporting the atoms along the nanofiber by mutually detuning the two counter-propagating fields which form the standing wave. The performance and limitations of the nanofiber-based transport are evaluated and possible applications are discussed.
\end{abstract}%

\section{Introduction}
The transport of cold and ultra-cold neutral atoms using time-dependent trapping potentials has been actively studied in recent years~\cite{Greiner01,Haensel01b,Kuhr01,Gustavson01,Miroshnychenko03,Bergamini04,Guenther05a}. From a technical point of view, a precise active control of the position of atoms~\cite{Dotsenko05} is a central requirement, e.g., for the implementation of an atomic ``quantum bus''~\cite{Kuhr03,Beugnon07}, for the deterministic and adjustable coupling of atoms to functional structures, such as optical resonators~\cite{Sauer04,Nussmann05,Teper06,Khudaverdyan08}, for optical clock applications~\cite{Middelmann11}, and for cold-atom-based microscopy~\cite{Gierling11}. 

Here, we realize the transport of a laser-cooled ensemble of cesium atoms using the two-color nanofiber trap demonstrated in Ref.~\cite{Vetsch10}. Using light-induced potentials, the atoms are trapped in the evanescent part of guided laser fields propagating through a subwavelength-diameter silica fiber \cite{LeKien04}. For a sufficiently small dia\-meter of the fiber, the latter only guides the ${\rm HE}_{11}$ optical mode~\cite{LeKien04b}, which can be laterally confined to a cross-sectional area on the order of $\lambda^2$, where $\lambda$ designates the wavelength. In contrast to optical conveyor belts based on focused free-beam optical dipole traps, our transportation scheme has the potential to overcome the diffraction-related limitation of the maximum transportation distance by exploiting the translational invariance of the trapping potential along the entire length of the nanofiber.

\section{Set-up}
\begin{figure}
	\includegraphics[width=8.3cm]{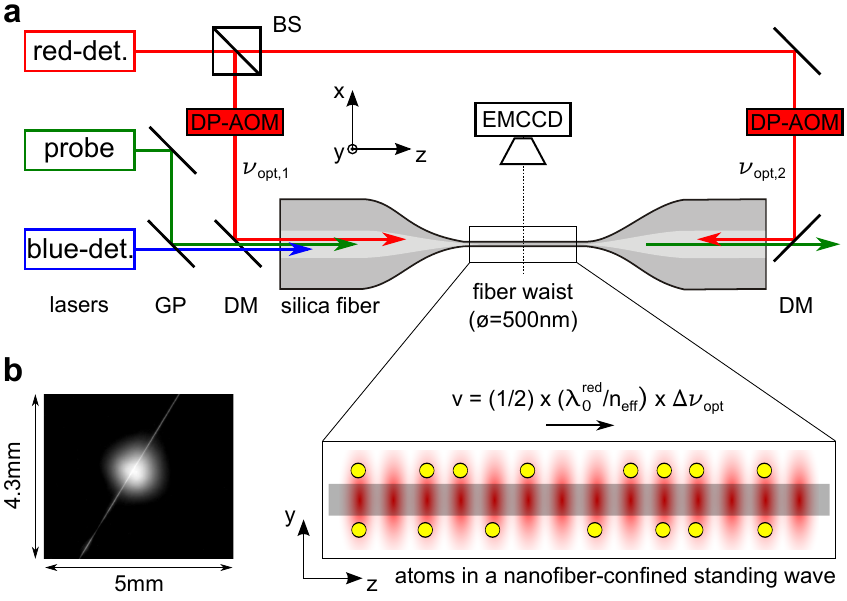}
	\caption{{\bf a} Schematic of the experimental set-up including the tapered optical fiber, the trapping laser beams, the probe laser beam, and an electron-multiplying CCD camera used for imaging the atomic fluorescence. BS: beam splitter, DM: dichroic mirror, GP: glass plate, DP-AOM: acousto-optic modulator in double-pass configuration. The atoms in the nanofiber trap and the evanescent standing wave pattern formed by the red-detuned trapping field are shown in the zoomed inset. {\bf b} CCD image of fluorescence of the atom cloud in the MOT which is overlapped with the nanofiber waist of the tapered optical fiber. The latter is also visible due to scattering of light from the MOT laser beams.}
	\label{fig:setup} 
\end{figure}

The experimental set-up of the nanofiber-based optical conveyor belt is sketched in Fig.~\ref{fig:setup}{\bf a}. With the exception of the added acousto-optic modulators (AOMs), it corresponds to the one introduced in~\cite{Vetsch10}. A tapered optical fiber with a 500-nm diameter nanofiber waist of $\unit[5]{mm}$ length is located in a vacuum chamber with a typical background gas pressure of $\unit[8\times10^{-10}]{mbar}$. Two far off-resonant dipole trapping laser fields as well as a near-resonant probe beam can be launched through the nanofiber. The relevant parameters of these fields are summarized in Table~\ref{tab:lasers}. A magneto-optical trap (MOT) serves as the source of cold atoms to load the nanofiber-based trap. It is overlapped with the nanofiber and is loaded from cesium background gas. Despite the presence of the nanofiber, the MOT performs normally~\cite{Russell12}. An image of the combined MOT--nanofiber system, re\-corded with a charge-coupled device (CCD) camera, is shown in Fig.~\ref{fig:setup}{\bf b}. The fluorescence of the atoms trapped in the MOT and the light of the MOT lasers scattered by the nanofiber are clearly visible.

Our nanofiber trap has been characterized in more detail in Refs.~\cite{Vetsch10,Vetsch12ArXiv}. The trapping potential consists of two arrays of local potential minima, located in the $y$-$z$-plane (see Fig.~\ref{fig:setup}{\bf a}). Typical trap frequencies are $(\nu_r, \; \nu_z, \; \nu_\varphi) = \unit[(200, 315, 140)]{kHz}$ in the radial, axial and azimuthal direction, respectively. The minima have a distance of $\unit[230]{nm}$ to the nanofiber surface, the trap depth $U_0/k_{\rm B}$ is about $\unit[400]{\mu K}$, where $k_{\rm B}$ is Boltzmann's constant, and we typically trap around 2000 atoms in total. The individual trapping sites are populated with at most one atom per site due to the collisional blockade effect~\cite{Schlosser02}, and the average filling factor is $\leq 1/2$~\cite{Vetsch12ArXiv}. A sketch of the nanofiber trap containing atoms is shown in the zoomed inset of Fig.~\ref{fig:setup}{\bf a}. For the periodic axial confinement of the atoms as well as their position control, the red-detuned laser beam is split, sent through AOMs operated in double-pass configuration and launched into the tapered optical fiber from both sides. The position of the resulting standing wave can then be controlled via the relative phase of the two counter-propagating red-detuned trapping laser fields. In particular, the standing-wave pattern can be translated at a constant velocity along the nanofiber by setting the mutual detuning of the two fields to a fixed non-zero value using the AOMs.
\begin{table}
	\begin{center}
		\begin{tabular}{|l||c|c|c|c|}
			\hline
			&  $\lambda_0$ & $P$ & pol.~plane & prop. \\
			\hline \hline
			red-det. & $\unit[1064]{nm}$ & 2 $\times$ $\unit[2.2]{mW}$ & $y$-$z$ & SW\\ \hline
			blue-det. & $\unit[780]{nm}$ & $\unit[25]{mW}$ & $x$-$z$ & RW\\ \hline
			probe & $\unit[852]{nm}$ & $\unit[500]{pW}$ & $y$-$z$ & RW\\ \hline
		\end{tabular}  
		\caption{Overview of the main laser field parameters. $\lambda_0$: free-space wavelength, $P$: power, SW: standing wave, RW: running wave. All fields are quasi-linearly polarized HE$_{11}$ modes with a non-vanishing longitudinal component. The predominant polarization components lie in the $x$-$z$-plane (blue-detuned trapping field) and $y$-$z$-plane (red-detuned trapping field and probe field), respectively, with the axes specified in Fig.~\ref{fig:setup}{\bf a}. }
		\label{tab:lasers}
	\end{center} 
\end{table}

\section{Measurement and Discussion}
The loading of atoms from the MOT into the fiber-trap is performed similarly to Ref.~\cite{Vetsch10}: After a MOT-loading period of $\unit[2]{s}$, the MOT magnetic field is switched off and the MOT cooling and re-pump lasers are switched to a molasses configuration, i.e., their powers are reduced and the red-detuning of the cooling light with respect to the free-space D2 $(F=4 \rightarrow F^\prime=5)$ transition is increased from $\unit[-15]{MHz}$ to $\unit[-80]{MHz}$. After this molasses phase of 100~ms duration, both the cooling and re-pump laser are switched off. The trapping laser fields are on during the entire experimental sequence and the counter-propagating far red-detuned fields are set to a mutual detuning $\Delta \nu_{\rm opt}=\nu_{\rm opt,1}-\nu_{\rm opt,2}$.

Due to the double-pass configuration of the AOMs, we have $\Delta \nu_{\rm opt}=2\cdot\Delta \nu_{\rm RF} = 2 \cdot (\nu_{\rm RF,1}-\nu_{\rm RF,2})$ where $\nu_{\rm RF,1}$ and $\nu_{\rm RF,2}$ are the frequencies of the RF signals applied to the AOMs (see Fig.~\ref{fig:setup}{\bf a}). A detuning $\Delta \nu_{\rm opt}$ of the two counter-propagating red-detuned fields results in a traveling fiber-bound lattice potential, moving at a velocity
\begin{eqnarray}
v = \frac{1}{2} \; \frac{\lambda_0^{\rm red}}{n_{\rm eff}} \; \Delta \nu_{\rm opt}
\label{eq:velocity}
\end{eqnarray}
along the z-axis, where $n_{\rm eff}$ is the effective refractive index experienced by the red-detuned trapping light propagating through the nanofiber~\cite{Yariv07} and $\lambda_0^{\rm red}$ is its free-space wavelength. In order to simplify the set-up and the experimental sequence, $\Delta \nu_{\rm opt}$ and thus $v$ are set to fixed values during the entire experimental cycle. If $v$ is comparable to or smaller than the thermal velocity of the atoms in the MOT cloud, the atoms are still transferred into the nanofiber-based trap during the loading sequence. In this case, as will be shown below, the transfer takes place during a short time interval around $t\approx t_0 + 77$~ms, where $t_0$ marks the start of ramping the MOT fields to a molasses configuration.

\begin{figure}
	\includegraphics[width=8.3cm]{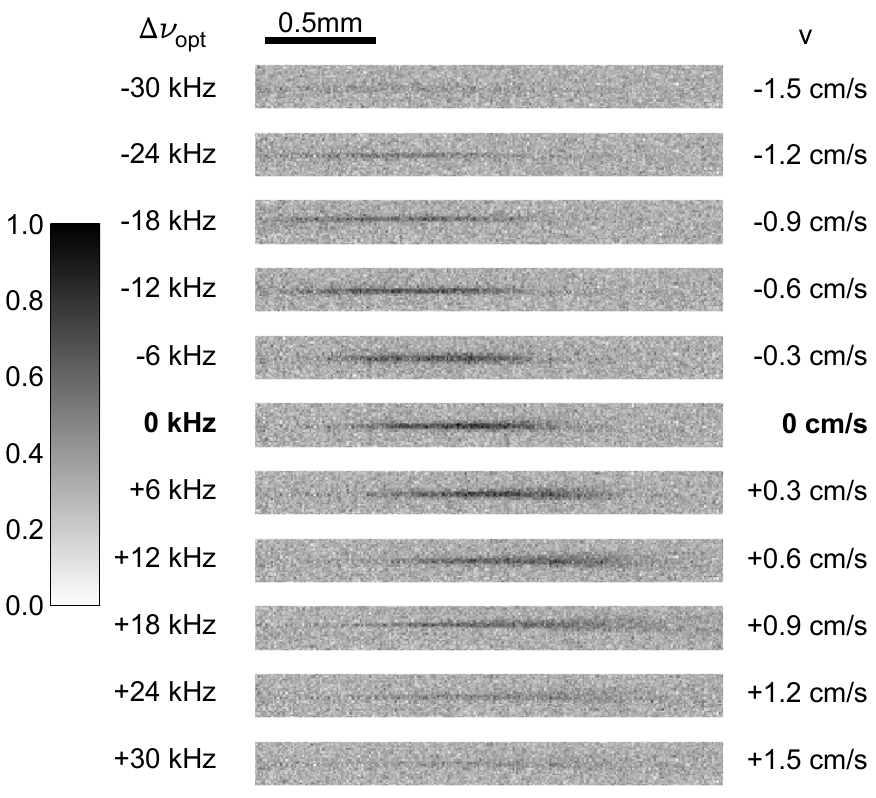}
	\caption{Fluorescence images of nanofiber-trapped atomic ensembles for different relative detunings $\Delta \nu_{\rm opt}$ between the two red-detuned, counter-propagating trapping laser fields. The corresponding velocities of the conveyor belt $v$ are also indicated. For an increasing absolute value of $\Delta \nu_{\rm opt}$, the atoms are transported over larger distances. For $|\Delta \nu_{\rm opt}|$ exceeding 18~kHz, the transfer of atoms from the MOT into the moving nanofiber-based trap becomes less efficient, as is apparent from a reduced fluorescence signal. For the maximum detunings of $\pm \unit[30]{kHz}$, the atomic ensemble is hardly discernible. The grey scale indicates the relative intensity of the fluorescence collected from the trapped atomic ensemble normalized to the maximum value of the $\Delta \nu_{\rm opt}=\unit[0]{kHz}$ measurement.}
\label{fig:transport}
\end{figure}

At time $t= t_0 + 108$~ms, we then record an in-situ fluorescence image of the nanofiber-trapped atoms with an electron-multiplying CCD (EMCCD) camera. The images for different values of $\Delta\nu_{\rm opt}$ are shown in Fig.~\ref{fig:transport} and have been recorded as follows: The atoms are excited with quasi-linearly polarized nanofiber-guided probe light at a red-detuning of $-20$~MHz with respect to the AC-Stark-shifted D2 $(F=4 \rightarrow F^\prime=5)$ transition frequency. The polarization of the probe light lies in the $y$-$z$-plane, such that the dipole emission characteristics yields the maximum signal in the imaging direction. To achieve a good signal-to-noise ratio, the images are the sum of 320 background-corrected exposures from consecutive experimental runs. As is clearly visible in Fig.~\ref{fig:transport}, the fiber-bound atomic ensemble can be transported along the nanofiber in both directions over distances of up to $\unit[0.3]{mm}$.

\begin{figure}
	\includegraphics[width=8.3cm]{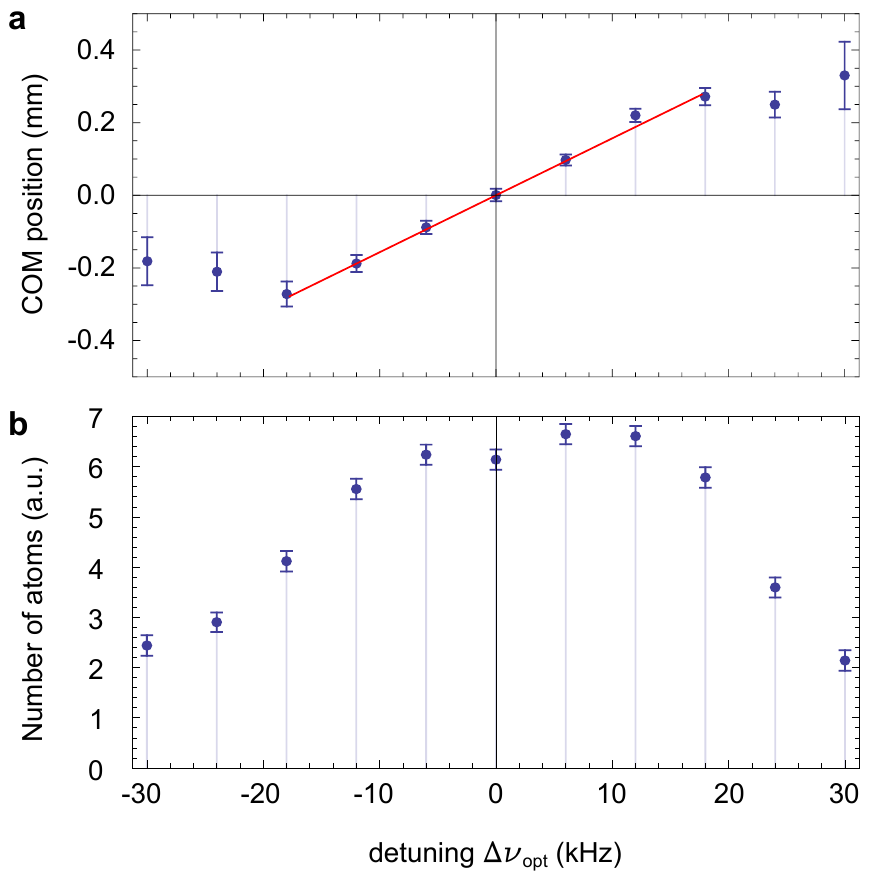}
	\caption{Quantitative characterization of the performance of the nanofiber-based conveyor belt. {\bf a} Center-of-mass position of the atomic ensemble as a function of the detuning $\Delta \nu_{\rm opt}$ of the two red-detuned trapping laser beams. For $|\Delta \nu_{\rm opt}| \lesssim \unit[18]{kHz}$ the COM position shifts linearly as is apparent from the linear fit (red solid line, see text). The statistical error bars increase towards larger detunings because of the reduced signal-to-noise ratio of the fluorescence images (see Fig.~\ref{fig:transport}). {\bf b} Number of atoms after transport as a function of detuning. A reduction for increasing $|\Delta \nu_{\rm opt}|$ is observed.}
	\label{fig:AtomsAndPos} 
\end{figure}

We now investigate the nanofiber-based transport more quantitatively by determining the center-of-mass \\(COM) position of the ensemble as well as the relative number of atoms in the trap from the fluorescence images. In Fig.~\ref{fig:AtomsAndPos}{\bf a}, the shift of the COM is plotted versus the detuning of the two counter-propagating trapping fields. The COM position is determined from the fluorescence signal in each image. Since only the motion along the fiber is of interest, we integrate the image along the direction orthogonal to the fiber. In order to determine both, the COM position of the atomic ensemble and its uncertainty, groups of $N=16$ adjacent pixels are binned. The groups are numbered from left to right by the index $j$. For each group, we compute the sum of the pixel brightness values, $B_{j}=\sum_{i=1}^{N} b_{i,j}$, as well as the error of this sum, $\Delta B_{j}$. The latter is estimated by computing the standard deviation of the $b_{i,j}$ values in each group $j$ and by multiplying it with $\sqrt{N}$. The COM position is computed by the weighted sum of the bin positions according to $\langle j \rangle=\sum_{j} j \cdot B_{j} / \sum_{j} B_{j}$. The error of each COM position is calculated by applying Gaussian error propagation to this formula using the $\Delta B_{j}$ values. In order to check the validity of this approach we verified that the size of the errors is largely independent of the bin size for $2 \leq N \leq 32$. Both, $\langle j \rangle$ and its error, are converted into real-space quantities using a calibration obtained in an independent measurement. 

For absolute detunings $|\Delta \nu_{\rm opt}| \leq \unit[18]{kHz}$, the COM shifts linearly with the detuning, indicating that the ensemble is transported in the conveyor belt for a fixed duration. Using Eq.~(\ref{eq:velocity}), the duration $\Delta t$ of the transport can be obtained by fitting $v\cdot \Delta t$ to the data in the range from $-18$~kHz to $18$~kHz in Fig.~\ref{fig:AtomsAndPos}{\bf a} yielding $\Delta t=31 \pm$1~ms. This allows us to conclude that the atoms are loaded into the nanofiber-based trap approximately 77~ms after the start of the loading sequence.

The data for $|\Delta \nu_{\rm opt}| > \unit[18]{kHz}$ deviate from the linear behavior. Moreover, the number of detected atoms after the transport also decreases in this detuning range, see Fig.~\ref{fig:AtomsAndPos}{\bf b}. This effect can be qualitatively understood by considering that the atoms are cooled by the molasses in the laboratory frame while the optical conveyor belt moves at a constant velocity with respect to the latter. If we assume that atoms which move in the same direction and with the same velocity as the moving conveyor belt are loaded most efficiently into the trapping sites, one would therefore expect the loading efficiency to be proportional to the fraction of molasses atoms with velocity $v$ given by Eq.~(\ref{eq:velocity}). This fraction is given by a one-dimensional Maxwell-Boltzmann distribution, i.e., a Gaussian as a function of $v$ with 1/$\sqrt{e}$-halfwidth of $\sigma_v =\sqrt{k_{\rm B}T_{\rm M}/m}$, where $T_{\rm M}$ is the temperature of the molasses-cooled cloud of atoms and $m$ is the mass of a cesium atom. Consequently, one expects the transfer efficiency to drop if $|v|$ exceeds $\sigma_v $ or, using Eq.~(\ref{eq:velocity}), if the detuning $|\Delta \nu_{\rm opt}|$ exceeds 
\begin{eqnarray}\label{eq:width_of_transfer}
2\sigma_v n_{\rm eff}/\lambda_0^{\rm red}\approx 16~{\rm kHz}\cdot\sqrt{T_{\rm M}}~, 
\end{eqnarray}
where $T_{\rm M}$ is in $\mu$K. In a separate measurement we record the expansion of our molasses-cooled cloud with a CCD as a function of expansion time and derive a temperature of $T_{\rm M}=(2.3 \pm 0.3)\mu$K. Inserting this temperature into Eq.~(\ref{eq:width_of_transfer}) yields a detuning range of $|\Delta \nu_{\rm opt}|\lesssim 24$~kHz over which the transfer efficiency is significant. This range is in reasonable agreement with the data in Fig.~\ref{fig:AtomsAndPos}{\bf b}.

In comparison to the experiments presented in \cite{Vetsch10}, the storage time of the atoms in the trap was significantly shorter here ($\tau^\prime \approx12$~ms vs.~$\tau \approx50$~ms). We attribute this reduced lifetime to an increased relative phase noise of the two counter-propagating lattice laser fields which leads to a position fluctuation of the lattice along the nanofiber, resulting in resonant heating of the atoms. The increase in phase noise is caused by the inclusion of the two AOMs into the beam path of the red-detuned trapping fields, see Fig.~\ref{fig:setup}{\bf a}, which are driven by two RF sources (Agilent MXG N5182A \& Agilent N9310A, where the first provided the 10~MHz external clock reference for the second). We measure the one-sided power spectrum $S_{\phi_{\rm RF}}(\nu)$ of relative phase fluctuations between the two RF signals~\cite{Savard97} and obtain $S_{\phi_{\rm RF}}(\nu_z)\approx 10^{-11}~$rad$^2$/Hz at the axial oscillation frequency of the trap $\nu_z=315$~kHz. Both red-detuned trapping fields pass through AOMs in double-pass configuration. Consequently, the relative phase noise of the RF signals powering the AOMs will be imparted on the red-detuned trapping fields, resulting in relative phase fluctuations between the two optical fields with a one-sided power spectrum of
\begin{eqnarray}
S_{\phi_{\rm opt}}(\nu)=4\cdot S_{\phi_{\rm RF}}(\nu)~. 
\end{eqnarray}
The axial position of the standing wave pattern formed by the two red-detuned counter-propagating trapping fields will fluctuate in the presence of relative phase fluctuations of the optical signals. The corresponding one-sided power spectrum $S_z(\nu)$ of the axial position fluctuations of the standing wave is given by
\begin{eqnarray}
S_z(\nu)=\frac{1}{4} \cdot \left( \frac{\lambda_0^{\rm red}}{2\pi n_{\rm eff}} \right)^2 \cdot S_{\phi_{\rm opt}}(\nu)~. 
\end{eqnarray}
According to Ref.~\cite{Savard97}, the resonant heating rate for atoms in a trap due to position fluctuations of the trap minimum can be calculated according to
\begin{eqnarray}
\langle \dot E \rangle &= 4 \pi^4 m \nu_z^4 S_z(\nu_z)~,
\label{eq:HeatingRate}
\end{eqnarray}
yielding $\langle \dot E \rangle \approx k_{\rm B} \cdot \unit[15]{mK/s}$ for our experimental values. In the absence of other heating mechanisms, this heating rate would lead to a trap lifetime of $\tau_{\phi} \approx (U_0 - k_{\rm B}T_0) / \langle \dot E \rangle \approx \unit[24]{ms}$, where $T_0\approx30~\mu$K is the initial temperature of the atoms in the nanofiber trap \cite{Vetsch12ArXiv}. Taking into account the trap lifetime without AOMs of $\tau \approx 50$~ms, the theoretically expected trap lifetime in the conveyor belt is thus given by 
\begin{eqnarray}
\tau^\prime=\left(\tau_{\phi}^{-1}+\tau^{-1}\right)^{-1}\approx 16~{\rm ms}~,
\end{eqnarray}
in reasonable agreement with our observation. Using a low phase noise dual-frequency RF source to drive the AOMs, it should be possible to overcome the current phase noise limitation of the trap lifetime. 

In principle, it should become possible to transport atoms over significant distances. For this purpose, one would load the atoms into a stationary standing wave which is subsequently accelerated. In the frame moving with the lattice, such an acceleration leads to a tilt of the potential. Assuming that the atoms are initially at a temperature much lower than the trap depth, a maximum acceleration of 
\begin{eqnarray}
a_{\rm max}=\frac{U_0}{m}\cdot \frac{2\pi n_{\rm eff}}{\lambda_0^{\rm red}} \approx \unit[1.6 \cdot 10^5]{m/s^2} 
\end{eqnarray}
can be applied before the trapping minima vanish~\cite{Schrader01}. Assuming this maximum acceleration, the maximum distance over which the atoms can be transported is given by $s=(a_{\rm max}/2)\cdot t_{\rm transp}^2\approx 8~{\rm cm}\cdot t_{\rm transp}^2$, where the transportation time $t_{\rm transp}$ is in ms. However, it should be noted that a limitation of $t_{\rm transp}$ might arise due to possible axial variations of the trapping potential due to imperfections of the nanofiber (inhomogeneous diameter, surface roughness, bulk scatters etc.). By transporting the atoms along the nanofiber, such variations would translate into temporal fluctuations of the trapping potential which could then lead to a velocity-dependent heating of the atoms in the conveyor belt. Moreover, for long accelerations, the required detuning between the RF drive frequencies might eventually exceed the bandwidth of the AOMs. And finally, the losses of the nanofiber section have to be small enough to guarantee a high contrast of the standing wave spatial intensity modulation over the entire length of the nanofiber. With our fiber pulling rig described in \cite{Warken08}, nanofibers with a waist length of up to $\unit[1]{cm}$ and a transmission of 98.9\%, measured with the method described in \cite{Wuttke12} and compatible with the above requirement, have already been manufactured successfully.

\section{Conclusion}
We demonstrated optical transport of cesium atoms over millimeter-scale distances using a nanofiber-based optical conveyor belt. In combination with the strong confinement of both the atoms and the trapping light fields, this technique might be used as a tool for coupling laser-cooled atoms in a deterministic way to functional devices like, e.g., whispering-gallery-mode resonators~\cite{Vernooy98b,Aoki06,Poellinger09} and solid state quantum circuits \cite{Hafezi12}. The transported atoms might also be used as a scanning probe for external fields similar to what has been realized with NV diamond nanocrystals \cite{Balasubramanian08}. Moreover, the transport should enable a sequential loading procedure of the nanofiber-based trap. In this way, larger ensembles of fiber-coupled atoms with optical densities on the order of a few hundred could be trapped. By sequentially loading two ensembles at a given distance with respect to each other along the nanofiber, this would also allow one to realize the recently proposed optical nanofiber-resonator with atomic mirrors \cite{Chang12}. Finally, it should be pointed out that due to very strong localization of the atoms in the nanofiber trap, the conveyor belt could be used to create a flux of cold atoms with very high brightness. This might facilitate nanofiber-based atom lithography as well as cold collision experiments with a high degree of control over the collision parameters.

\begin{acknowledgement}
This work was supported by the Volks\-wagen Foundation (Lichtenberg Professorship), the European Science Foundation (EURYI Award), and the Austrian Science Fund (CoQuS Graduate school, project W1210-N16). The authors wish to thank LOT-Oriel for the loan of the EMCCD camera.
\end{acknowledgement}

% ------------------------------------------------------------------------
%Steps to correctly include the bibliography:
%1: Write refs. in bibtex
%2: Uncomment \bibliographystyle and \bibliography lines below and Ctrl-Shift-L, Ctrl-Shift-B. Ctrl-Shift-L, Ctrl-Shift-L
%3: Comment the lines again.
%4: To be sure, copy and paste the contents of the Paper.bbl file below i.e. don't use Bibtex.
%5: Ctrl-Shift-X to check the output
%6: Repeat again for new references.
%7: Use Bibliography_custom.bst style file because it has been edited by me to have the correct format (I think)
%8: Don't forget to kill .aux and .bak files if necessary

%\bibliographystyle{Bibliography_custom} % Bibliography style file
%\bibliography{transportR1} % Bibliography database file .bib

% ------------------------------------------------------------------------
\end{document}